\documentclass[12pt]{article}
\usepackage{graphicx}
\usepackage{euscript}
\usepackage{amsmath}
\usepackage{amsthm}
\usepackage{amssymb}
\usepackage{natbib}
\begin{document}
\begin{center}
\large 
\bf
The Age-Specific Force of Natural Selection 
\\
and Walls of Death
\\
\rm
\vspace{.2in}
Kenneth W. Wachter, Steven N. Evans, and 
\\
David R. Steinsaltz 
\\
30 June 2008
\\
\end{center}
\normalsize 
\rm
\begin{abstract}
W. D. Hamilton's celebrated formula for the 
age-specific force of natural selection 
furnishes predictions for senescent
mortality due to mutation accumulation,
at the price of reliance on a linear approximation.
Applying to Hamilton's setting the full
non-linear demographic model for mutation 
accumulation of \cite{diploid},
we find surprising differences.  
Non-linear interactions cause the collapse
of Hamilton-style predictions in the most 
commonly studied case,  refine predictions
in other cases,  and allow Walls of Death at
ages before the end of reproduction.   
Haldane's Principle for genetic load 
has an exact but unfamiliar generalization.  
\end{abstract}

\it
Acknowledgments:
\rm
This work has been supported by Grant AG-P01-008454 from
the National Institute on Aging, and the research of 
SNE has been supported in part by Grant DMS-0405778 
from the National Science Foundation.   
We thank Shripad Tuljapurkar and Nicholas Barton 
for insightful suggestions.

\section{
The Force of Selection}
\label{sec.force}

\par\indent

The best-known formula at the intersection
of genetics and demography is doubtless
W. D. Hamilton's ``age-specific force of
natural selection''.  
\cite{wH66}  differentiated a measure of fitness,
Lotka's intrinsic rate of natural increase,
with respect to an increment to age-specific mortality 
at an age $a$.   Thus he obtained a linear approximation
for loss in fitness due to any deleterious
mutations that raised mortality at
any one specific age.  The greater the loss in
fitness,   the faster should mutant alleles be
selected out of a population and the fewer should
be found at equilibrium as recurring mutations
balance natural selection.

By this route, Sir Peter Medawar's concept of
mutation accumulation as an evolutionary reason
for senescence takes on mathematical form.  
As in \cite{pM52},   \cite{cF90}, 
or \cite{bC00a},
the idea involves genetic load produced
by large numbers of mildly deleterious mutations 
occurring at widely separated loci, each with
some small age-specific effect on vital schedules.
As developed by Brian
\cite{bC94}
this framework guides the interpretation of many
experiments in aging research and population genetics.  
A recent perspective is offered by \cite{FP07}.  
For definitions, see Section \ref{sec.linear}. 

Hamilton's work, reprinted in \cite{wH95}, has
been assessed and extended by \cite{aB08}.  
Sophisticated genetic models of mutation-selection balance
are covered in an authoritative book by \cite{rB00}.  
Demographers mainly put up with less sophisticated
models of the genome in return for more refined
treatments of age-specific structure, as we do here.  
Age-specific predictions for vital schedules are
sometimes robust to details of genetic specification,
in line with a principle of \cite{jH37}   
which equates the population loss in fitness from
genetic load to the total mutation rate,
independent of the form of action of mutations.
Current interest has been
stimulated by the expansion of biodemography,
reviewed in \cite{WF97},  \cite{jV98}, \cite{jC03}, and \cite{CT03}. 
%

The theory built on Hamilton's formula is, in short,
a centerpiece for demographic research.  
But Hamilton's formula fails to be self-consistent.
Its reliance on linear approximation requires 
total increments to age-specific mortality to stay 
small where the formula predicts them to grow large.

Non-linearity is built into the fitness measure.  
As deleterious mutations arise, their overall effect 
differs from the sum of their individual effects.  
Diminished survival at any one reproductive age 
necessarily leaves less reproduction to be lost 
by a drop in survival at any other reproductive age.  
This interaction,  a key feature of mutation accumulation,
is set to zero in the linear framework.  
If all effects were small enough, little accuracy would
be lost,  but the prediction turns out to be for large
cumulative effects from small individual ones.
Richly suggestive as it is,  Hamilton's formula can only
be taken so far and no further.  How far is too far?
The answer has remained in doubt.

In \cite{diploid}
we developed a full non-linear age-specific model
for mutation accumulation.  
In this paper we apply our model to Hamilton's
setting,  in which the effect on age-specific 
mortality of each deleterious mutation is concentrated 
at a single age.  As explained shortly, these
effects are called ``point-mass increments''
and they are added onto the continuous-age version  
of mortality rates called the ``hazard function''.  
This specification is not a realistic one 
for actual mutations,  but it is of central 
interest given its dominant role in prior work.   
A companion paper in preparation
applies the new model to cases with more plausible
specifications of distributed mutational effects.  
With point-mass increments, we are fortunate
to obtain closed-form expressions, a particular
advantage since numerical simulations may not
readily reveal whether finite solutions do or not exist.

The most studied case with point-mass increments
was put forward by \cite{bC01}.  
He showed that the widely-observed Gompertz-Makeham
form for age-specific mortality would be predicted
exactly by the linear approximate model if one took 
mutation rates, background hazards, and fertility all 
constant across ages beyond an age of maturity.  
The Gompertz-Makeham form has a hazard function 
rising exponentially with age added onto a constant 
background level.  Charlesworth was able to 
generalize the Gompertz-Makeham prediction beyond 
the point-mass setting.   In his framework, a ``Wall of Death'' 
with infinite hazards and zero survivorship  
could occur,  at but not before an upper age limit
to fertility, if such an age limit were imposed. 
Charlesworth's elegant results with the linear
approximate model have shaped thinking in biodemography.

Our paper presents three main findings for the
full non-linear model with point-mass increments,  
all of them surprising:
\begin{enumerate}
\item
In the elementary case with constant rates,
the linear model breaks down when 
non-linear interactions are taken into 
account.  An equilibrium ceases to exist,
and accumulating mutations drive survival to 
zero at every adult age.
\item
When provisions are introduced that preserve
survival at some adult ages, 
a ``Wall of Death'' can occur before rather 
than at the oldest age of reproduction.
\item
A generalization of Haldane's Principle holds
exactly 
but it takes an unfamiliar form.
\end{enumerate}

These results demonstrate the limitations of the
traditional approach grounded in Hamilton's
formula.   They furnish guidance 
for constructing more realistic specifications 
of mutational action.   It is especially
revealing to learn that Gompertz-Makeham 
hazard functions cannot withstand certain
simple kinds of sustained mutational pressure
when interactions are taken into account.  
In the interpretation which we shall offer,
equilibrium solutions disappear because
they collapse to something outside the
model specification.  The ``missing''
equilibrium is a stylized life history  
in which all fertility is concentrated
in a burst at a single age, followed by
immediate death,  a life history 
which \cite{sT97} 
calls ``the salmon limit''.  

Like the linear models,  our non-linear model is
an infinite-population model in continuous time.   
The representation of the genetic structure is
kept somewhat stylized in order to allow 
ramification of the demographic structure.
The model follows in the tradition of \cite{KM66}.  
Inheritance is diploid, with random mating,
weak selection, and fitness calculated for
individuals rather than for mating pairs.
At each of a large or infinite number of sites is
found either a wild-type or a dominant 
deleterious mutant allele.   Alleles with the
same effect on the hazard function are treated
as if they were copies of the same allele,
though found at different sites.
Back mutation is taken to be negligible.
A randomly selected member of the population carries 
some collection of such mutant alleles,  and the state
of the population is described by a joint
probability distribution for the counts of
alleles of different kinds, that is, of
alleles acting at different ages.
The linear models posit Poisson distributions,
consistent with a derivation found on page 137 of \cite{rD02}.
Our model recovers this Poisson property for
solutions, in a sense explained in Section \ref{sec.nlmodel}.
The property follows from an assumption 
called ``Free Recombination,'' spelled out in \cite{diploid}. 

Genetic recombination makes no difference when
interactions are all suppressed by linear approximations,
but it matters in non-linear models.  
Under Free Recombination,  recombination
is assumed to act on a more rapid timescale than 
mutation and selection,  erasing linkage disequilibrium
among sites involved in the mutation accumulation
process.  An alternative non-linear model without
recombination is presented in \cite{SEW05} but 
will not be treated in this paper.  

When age like time is taken to be continuous,
adding an increment to the hazard function ``at''
an age $a$ does not strictly make sense.  The remedy is
to think of adding a point mass or ``delta function''
of some size $\eta$ to the hazard function,  
equivalent to adding onto the cumulative hazard
function a step function with a single upward
step of size $\eta$ at age $a$.  Hamilton's
formula then follows by differentiating Lotka's $r$
with respect to $\eta$ applying implicit
differentiation to the Euler-Lotka Equation which
defines $r$.   
In the application of our model to Hamilton's
setting,  the mutational effects are point-mass
increments.  

Hamilton's formula applies in principle to favorable
as well as to deleterious mutations. However, outside
the context of mutation accumulation the formula is  
uninformative with respect to age-specific shapes
of vital schedules.  Any recurring favorable
mutation eventually goes to fixation.  
The rapidity of fixation does not matter to the
ultimate contribution.  Transient effects are
possible,  but those would depend on independent
definition of some origin for time. Parallel
studies of fertility are also of interest 
but outside our present scope.  
Here we concentrate
on deleterious mutations affecting survival
and their demographic consequences. 

\section{
The Linear Approximation}  
\label{sec.linear}

\par\indent

Our non-linear model is derived from dynamic 
equations in \cite{diploid} and may be regarded as a 
limiting form of standard discrete-population 
genetic models such as those of \cite{BT91}  and \cite{KJB02} 
in the asymptotic regime of weak selection and mutation.
Although the derivation is complicated,  the formulas 
for predicted hazards are simple.  They are
presented in Section \ref{sec.nlmodel}.  
An informal account of why the answers take the shape
they do is offered in this section through an examination of 
the linear approximation built into Hamilton's formula.

For the convenience of readers, we review
some demographic terminology.  Let $\zeta$ be a
random variable that represents the life-span
of an individual picked at random from a population,
or from a subpopulation sharing some attribute.
The {\em survivorship function} $l_x := \mathbb{P}\{\zeta > x\}$
is the probability of survival
from birth to age $x$.  The {\em cumulative hazard}
at age $x$ is $-\log l_x$. The
{\em hazard function} itself at age $x$ is
minus the derivative of the logarithm of the survivorship,
when the derivative exists.
The product of the survivorship function
$l_x$ with the {\em age-specific fertility
rate} $f_x$ is the {\em Net Maternity Function} $f_x l_x$.
Fertility for males is taken to be governed by the rates
for their female mates.  The models do not
explicitly segregate individuals by sex. 
The area under the Net Maternity
Function (that is, $\int_0^\infty f_x l_x \, dx$)
is the {\em Net Reproduction Ratio} or $NRR$.
The $NRR$ measures generational replacement, the
ratio of the size of the next generation to the current one.

If the survivorship function $l_x$ and fertility 
schedule $f_x$ do not change over time, then they 
lead to a population with an unchanging proportional
distribution of ages called a {\em stable population}.
The growth rate of the stable population (the slope of
the logarithm of population size over time) is
{\em Lotka's intrinsic rate of natural increase} $r$.
Lotka's $r$  is  the unique real root of the Euler-Lotka
Equation $ 1 = \int \, e^{-rx} \, f_x \, l_x \, dx $.
That is, $r$ is the unique real zero of the
logarithm of the Laplace
Transform of the Net Maternity Function.  On
evolutionary timescales, the parameter
$r$ is assumed to have
been close to zero,  the growth rate of
a {\em stationary population}.

Hamilton's  expression for the force of natural 
selection is a function of the age $a$ at which
the effect of a mutant allele is assumed to be
concentrated.  It is calculated from background
or extrinsic schedules for $l_x$ and $f_x$,
called the  {\em baseline} schedules.
If time is measured in units of generations, 
with generation length equal to the stable population 
mean age at childbearing,  and if Lotka's parameter $r$ 
is set to the stationary level $ r = 0$, 
the force of selection reduces to the simple
form $w(a) = \int_a^{\infty} \, f_x \, l_x dx $, 
the expected number of offspring produced after
age $a$ by an individual chosen at random from
the population, offspring that would be lost by 
death at age $a$.
With time in units of generations, the expression
is the same whether the fitness measure being differentiated
is Lotka's $r$ or the Net Reproduction Ratio. 

As we have said, Hamilton's expression gives a 
linear approximation to the fitness cost of 
mutations which add point-mass increments 
of some size $\eta$ to the hazard at age $a$.  
The loss from $n$ such mutations is being approximated
by $ n \eta $ times $ w(a) $, even though the fitness
itself depends exponentially on $n$ and $ \eta $.   
Deleterious mutations are taken to be occurring
at a rate $ \nu(a)$ and accumulating.
For outflow $n \eta w(a) $ due to selection to balance 
inflow $ \nu(a) $ due to mutation at equilibrium,
the linear model sets $ n \approx  \nu(a)/(\eta w(a) ) $.
The approximation for the increment to the hazard
function is then $  n \eta = ( \eta \nu) /( \eta w ) = \nu/w $,
independent of the size $ \eta $ of the effect.  
See \cite{bC01} for a full account.

A contradiction arises because mutational effects 
of tiny size imply equilibrium increments to 
the hazard function of hefty size.
Small values of $\eta$, which ought to help the approximation,
strain it by bolstering the equilibrium value
for $n$.  The force on any one of $n$ alleles acting at $a$
responds to alterations to the survivorship function due
to the other $n-1$ alleles, as well as to alterations 
due to alleles acting at other ages.  In other words,
interaction terms enter the calculation.  
For consistency, $w$ ought to be computed from a
new $l_x $, reflecting these alterations.   
But then we would have new increments to the hazard
function and need a new new $l_x$, over and over. 

The hazard function at equilbrium computed from the
survivorship function for the population as a whole
may be written as a sum of a contribution $ \lambda(x)$ 
due to baseline risks and increments $h(x)$ due 
to the mutant alleles distributed across the population.  
The linear approximation based on Hamilton's formula
puts
\begin{equation}
\label{E:hamha}
h(a) \approx  \frac{\nu(a)}{w(a)} = 
        \frac{\nu(a)} { \int_a^{\infty} \, 
   \exp \left( -\int_0^x \, \lambda(y) dy \right) \, f_x \, dx} 
\end{equation}

Since all increments affect selective costs, 
a guess at correction might be to replace the baseline hazard
with the total hazard, leading to an equation with $h$ on both sides:
\begin{equation}
\label{E:goodha}
h(a) = \frac{\nu(a)} { \int_a^{\infty} \, \exp \left( 
       -\int_0^x \, ( \lambda(y) + h(y)) \, dy \right) \, \, f_x \, dx} 
\end{equation}
This guess turns out to be the actual 
equation for $h$ with point-mass increments in the
full non-linear model.

The simplicity of Equation \eqref{E:goodha} hides some
subtleties.  The function $h$ in \eqref{E:goodha} is the
hazard function computed from the population survivorship
function.  Also called the ``aggregate''  hazard function,
it differs from the average hazard function.   
Across the population, the set of mutant 
alleles present in each individual is random.    
Different members carry different genetic loads.   
Without such heterogeneity,  natural selection would 
have nothing to select.   Members carrying heavier 
loads tend to die at younger ages. 
This process of ``demographic selection''  makes 
hazards observed among survivors lower than 
hazards evaluated by averaging over the distribution 
of alleles inherited at birth.  It is not obvious
whether formulas should feature the average hazard
or the aggregate hazard.  
As shown in Section \ref{sec.nlmodel},  the 
non-linear model picks out the aggregate hazard,
as a consequence of computing a statistical expectation
value for the marginal cost of each mutant allele.  

%

Another point concerns the specification of
fitness costs in the presence of heterogeneity.
When linear approximations are being used,  
it makes no difference whether costs are based on
Lotka's $r$ or on the Net Reproduction Ratio, the $NRR$.
Pages 136 to 146 of \cite{bC94} give a careful
examination of first-order and second-order terms.  
For non-linear models, the choice does make some
difference.   As \cite{bC00a} points out on page 930,  
the $NRR$ is the appropriate fitness measure for our
purposes.   The alleles
are not invading a population but are being
held at equilibrium frequencies.  
Measuring selective cost by reductions in 
the $NRR$ makes frequencies agree with classical 
formulas for single-locus models.

The demographic background to the specification
of selective costs may be clarified by
reference to stable population theory.  
The population members who carry a particular 
collection of mutant alleles make a contribution
to the next generation given by their 
mutation-dependent $NRR$.  Thanks to new mutations 
as well as to recombination, their offspring 
do not carry identical collections of alleles.  
Groups of carriers are broken up each generation,
before they establish their own special
stable age structure or their own special
values of Lotka's $r$, the growth rate
that occurs with a stable age structure.
At equilibrium all groups of carriers share
the same growth rate, since their numbers
are replenished by new mutations to balance
their loss in numbers due to natural selection.
For this reason,  the $NRR$ rather than $r$
determines selective costs for mutation accumulation.

We are now in a position to preview our
chief results. 
In the linear theory,  the force of selection $w(a)$
is non-zero for any age at which the net maternity
function calculated from the baseline schedules
is non-zero.  Non-zero $w(a)$ implies  
finite equilibrium numbers of
mutant alleles acting at each age $a$.    
Numbers may tend to infinity as $a$ approaches
the last age of reproduction, imposing a 
``Wall of Death'' at that age but not before.
Late-acting mutations have no impact on the
numbers of earlier-acting mutations.  
Selective pressure at an age abutting on
the Wall of Death is calculated as if there
were no Wall of Death.  

In the full non-linear theory,  in contrast,
a Wall of Death, by erasing all net reproduction 
beyond itself,  reduces selective pressure
against mutations acting at slightly younger
ages.  Reduced pressure may mean that
selection cannot balance the rate of new
mutations and equilibrium conditions cannot
be satisfied at a younger age.  The Wall of
Death at the older age comes to imply a Wall of 
Death already at a slightly younger age.
If this chain of implication proceeds 
unchecked through younger and younger ages,  
it may mean that no equilibrium solution with 
finite mean numbers of mutations
exists at all.  We say that ``the solution unravels'',
or, more precisely, that an attempt to solve for
the equilibrium distribution of mutant alleles 
in the population unravels.   
We show in  Section \ref{sec.unravel}
that such pathology can in fact occur.  

Unraveling can be prevented in some biologically
sensible ways.  But equilibrium mean numbers of
mutations can still go to infinity at ages before 
the last age of net reproduction,
as we show in Section \ref{sec.walls}.  
A Wall of Death may be found at earlier ages
than the linear framework permits. 

Mutation accumulation has the appealing property
that predictions are insensitive to some of the
details of specification.  Working with genetic  
models without age structure,  \cite{jH37}, page 341,
announced that
\begin{quote}
``... the loss of fitness to the species depends entirely
on the mutation rate and not at all on the effect of the
gene upon fitness of the individual carrying it ...''
\end{quote}
He found the sum total of mutation rates at different
sites to be approximately equal to the resulting decrement
in the logarithm of fitness, a measure of ``genetic
load'' essentially equivalent to our selective cost.  

Haldane worked with two alleles per site and imposed  
a strong assumption about independence of mutational effects.   
Generalizations to multi-allele models are proved
on pages 105 to 112 and 143 to 153 
of \cite{rB00},  and a version holds in the 
age-specific linear framework.  
In our non-linear model with point-mass increments,
the aggregate population hazard and the population loss
in fitness are strictly independent of the sizes of the
increments, as we show in Section \ref{sec.haldane}. 
The total mutation rate, however, is not equal to the
total selective cost,  but to a less conventional
function of the demographic schedules.

\section{
The Non-Linear Model} 
\label{sec.nlmodel}

\par\indent

We now present the non-linear model from \cite{diploid} 
and show how it leads to the equation \eqref{E:goodha}
for predicted aggregate hazards in the special case
of point-mass increments.  

Mutant alleles $m$,  distinguished by their age-specific
effects, are drawn from a space $\mathcal{M}$.  
In our point-mass setting, $m$ corresponds to an age 
of action or ``age of onset'',  a point on the real line,
and $\mathcal{M}$ is the positive real line itself.   
Each individual carries some finite batch of mutant 
alleles denoted by the letter $g$,  and we use the 
word ``genotype'' as shorthand to refer to it. 
A member who carries no mutant alleles is  
said to carry the ``null genotype'' $ g = 0 $, 
with wild-type alleles at every site.  

An individual sampled at random from the population
carries a random batch of points $G$.  The count 
of points of $G$ in any interval of $\mathcal{M}$ is a 
random variable.  The mean of this random variable 
is given by the area within the interval under 
a curve $\rho$ called the 
\it
intensity.
\rm
(Technically speaking, $\rho$ is a density with 
respect to Lebesgue measure on the line.)
In the non-linear model under the assumption of
Free Recombination,  \cite{diploid}  prove that the
random counts in disjoint intervals are independent
random variables with Poisson distributions.  
Such a 
\it
Poisson point process 
\rm
is uniquely determined by its intensity.
It can also be defined
for more general choices for $\mathcal{M}$ 
including sample spaces for stochastic processes.  
The points of the Poisson process are points of age, 
not points of time.  The Poisson property holds
at any given time and also at equilibrium,
if an equilibrium exists. Background on Poisson 
processes may be found in textbooks 
like \cite{oK02}, Chapter 12.    

Each application of the general model requires three ingredients:
the age-specific profiles for the actions of mutant alleles,
the rate at which new mutations enter the population,
and the selective cost which gradually drives mutant alleles
out of the population.  

Here each mutation profile is written as a
function $ \kappa (m,x) $ of the index $m$ and
an age variable $x$.   The function $\kappa(m,x)$
is multiplied by a size factor $\eta(m) $ and 
added onto the cumulative hazard function.  
In other words, the cumulative hazard function defined
for a subpopulation of individuals with genotype $g$ 
is formed by starting with the cumulative baseline hazard
and adding a term $ \eta(m) \kappa(m,x) $ for each 
point $m$ in the batch of points $g$.  In demographic language,
alleles act like independent competing risks in a multiple
decrement lifetable.  Other interesting forms of action 
including proportional hazards are studied by \cite{aB08}, Chapter 2.

In this paper, following Hamilton, 
we take the profile $\kappa(m,x)$
to be a step function with a unit step at the
age of onset for $m$, corresponding to a
point-mass increment to the hazard itself.  
Equations do not depend on this 
special choice for $\kappa$ until Equation \eqref{E:kapintegral}
at the start of Section \ref{sec.pointmass}.

The number of new mutations per generation with $m$
from some interval of $\mathcal{M}$ is given by the area 
within the interval under a curve $\nu(a)$ called the 
\it
mutation rate.  
\rm
Like the intensity $\rho $, the mutation rate is 
a density with respect to Lebesgue measure. 
As in the elementary cases studied by \cite{bC01},
we usually take $\nu(m)$ to be constant or nearly constant
across ages within the reproductive span  
because we are interested in structure that arises from 
the logic of natural selection rather than
from structure arbitrarily built into assumed mutation rates.
Concentrating on adult mortality, we make $\nu$ 
vanish for ages of onset below some age at maturity $ \alpha $
at which fertility and exogenous baseline mortality commence.  
The level of fertility is tuned to produce a stationary
population at equilibrium,  cancelling out effects of juvenile
mortality and letting us omit them here.

The selective cost function $S$ is a non-negative function of $g$,
here taken equal to the decrement in the $NRR$ due to the
mutant alleles included in $g$.  This choice was explained
in Section \ref{sec.linear}.   Formulas are given 
in Equations \eqref{E:Sofg}  and  \eqref{E:lxg} below.     

Since a Poisson point process is uniquely determined by
its intensity,  the theorems in \cite{diploid} allow 
the population over time to be described by an equation 
for the intensity $\rho_t(m) $ over time.   
This dynamic equation involves an expectation 
value, written $ \mathbb{E}_{\rho} $, 
which averages over the random genotypes $G$ 
of randomly selected members of the population,
using the intensity function for the Poisson process:
\begin{equation}
\label{E:rhot}
\frac{d \rho_t(m)}{dt} = \nu(m) -  \rho_t(m)  \mathbb{E}_{\rho_t} \,  
               \left[ S(G + \delta_m ) - S(G) \right] 
\end{equation}
The symbol $G + \delta_m$ denotes a genotype with all 
the mutant alleles in $G$ plus one
copy of $m$,  using $\delta_m$ to stand for the measure
which puts mass $1$ on the point $m$.   
Equation \eqref{E:rhot} sets change equal to inflow minus
outflow, with inflow given by the mutation rate and
outflow given by mean numbers times the marginal selective
cost of each additional mutant allele.
An equilibrium intensity $\rho$, if one exists, has the left-hand
side equal to zero,  requiring $\rho$ to satisfy 
\begin{equation}
\label{E:equil}
\nu(m) =  \rho(m) \;   \mathbb{E}_{\rho} \;  
            \left[ S(G + \delta_m ) - S(G) \right]  
\end{equation}

We calculate predicted hazard functions at equilibrium by
substituting demographically meaningful expressions 
for $S(g + \delta_m)$ and $S(g)$ for each fixed $g$.
The selective cost function $S$ measures the fitness 
of genotypes on an implicitly logarithmic scale.   
We continue to write $ f_x $ for the fixed baseline age-specific
fertility schedule and $ \lambda(x) $ for the baseline hazard
rate, so that the survivorship function for the null 
genotype is given by $ l_x(0) = \exp( -\int_0^x \, \lambda(a) da)$.

The function $S$ can be defined to equal 
\begin{equation}
\label{E:Sofg}
S(g) :=   \int \, f_x \, l_x(0) dx - \int \, f_x \, l_x(g) dx 
\end{equation} 
The integrals are all taken over ages for which the
integrands are non-zero.  
The function $l_x(g)$, the probability of survival to 
age $x $ for members with genotype $g$, is derived 
from hazards that include increments from mutations in $g$.
We add up the increments on top of the baseline to form the
cumulative hazard, insert a minus sign, and exponentiate
to recover the survivorship: 
\begin{equation}
\label{E:lxg}
l_x(g) := l_x(0) \, \exp \left(  
               - \sum_{m^{\prime} \in g} \eta(m^{\prime}) 
                  \,  \kappa(m^{\prime} ,x) \right) 
\end{equation}
The prime on $m$ is a reminder not to confuse the index of
summation with the mutant allele $m$ whose marginal cost we
seek to calculate. 

Adding a copy of some particular mutant allele $m$ 
to the alleles already in $g$ multiplies this survivorship 
by $ \exp( -\eta(m)  \kappa(m ,x)) $,  
with a marginal cost equal to \begin{equation}
\label{E:marcost}
S(g + \delta_m ) - S(g) =  \int_{\alpha}^{\infty} \,  
                ( 1 - e^{-\eta(m) \kappa(m,x)} )
                        f_x l_x(g) \, dx   
\end{equation}

The theory tells us that our non-linear counterpart
to Hamilton's age-specific force of natural selection 
is the expectation value of this marginal cost,
formed by letting $g$ range over randomly selected
genotypes $G$ from the Poisson point process.  
The integrand factors into a fixed part involving only the 
extra allele $m$ and a random part involving only $G$, 
namely the net maternity $ f_x l_x(G) $ for the 
subpopulation with genotype $g = G$.   
By Equation \eqref{E:lxg}, the net maternity function 
resembles a Laplace Transform for the distribution of $G$.
This Poisson Process expectation 
can be taken in closed form, 
as, for instance, on page 227 of \cite{oK02}: 
\begin{equation}
\label{E:Elxg}
 \mathbb{E}_{\rho} \, \left[ l_x(G) \right]  
       =  l_x(0) \, \exp \left(     
      - \int_{\mathcal{M}} ( 1- e^{-\eta(m^{\prime}) \kappa(m^{\prime}, x)}) \,  
          \rho(m^{\prime}) dm^{\prime}   \right )     
\end{equation}

\section{
Solutions with Point Mass Profiles}
\label{sec.pointmass}

\par\indent

In Section \ref{sec.nlmodel}, the expressions do 
not depend on the form of $\kappa$.   Now we specialize
to point-mass profiles. The step-function form for $\kappa$  
leads to further simplification.  We substitute from 
Equation \eqref{E:marcost} in the equilibrium formula,
Equation \eqref{E:equil},  replacing $m$ with its 
age of onset $a$.   
\begin{equation}
\label{E:kapintegral}
\nu(a) = \rho(a) \, \int_{\alpha}^{\infty} \, ( 1 - e^{-\eta(a) \kappa(a,x)} )
                       \,  \mathbb{E}_{\rho} \, [ \, f_x l_x(G) \,] \,  dx
\end{equation}

The  step function preceding the expectation value 
restricts the range of integration to $x>a$, 
replacing $\alpha $ by $a$,  since the
integrand vanishes for $x$ below the age of onset. 
The random factor also simplifies, since the integral
over $\mathcal{M}$ in Equation \eqref{E:Elxg} can be replaced by
an integral over ages from $\alpha$ to $x$.  It can 
be written in terms of the increment $h$ to the aggregate
hazard discussed at length in Section \ref{sec.linear}.    
The increment $h$ is defined by 
\begin{equation}
\label{E:hdef}
h(x) :=  - \frac{d}{dx} \log \left( 
               \mathbb{E_{\rho}} \, [ \, l_x(G)/l_x(0)\,] \, \right) 
\end{equation}
We can express the aggregate population survivorship function 
in terms of $h$:
\begin{equation}
\label{E:aggsurv}
\mathbb{E}_{\rho} \, [ \, l_x(G) \,] = 
  \exp \left( -\int_{\alpha}^{x}\bigl[\lambda(y) +h(y) \bigr]dy \right) ,
\end{equation}
We have decomposed the aggregate population hazard into 
independent competing risks $\lambda $ due to baseline 
and $ h$ due to genetic load.
When we substitute \eqref{E:aggsurv} in \eqref{E:kapintegral} 
and compare with \eqref{E:Elxg}, we see how $h$ depends
on $\rho$ and $\eta$:
\begin{equation}
\label{E:hofx}
h(y) :=  ( 1 - e^{-\eta(y)} ) \rho(y) 
\end{equation}
Furthermore, we see that Equation \eqref{E:kapintegral}
is equivalent to the expression introduced in Section \ref{sec.linear},
Equation \eqref{E:goodha}.  As promised, we have derived 
the equation for $h$ from the general equations
of the non-linear model.  

We now specialize our choices of baseline schedules 
for the sake of our main applications.   
We choose a value for $\alpha$, the age of maturity 
below which exogenous baseline mortality and fertility
are taken to vanish.  Beyond $\alpha$, the baseline hazard
is taken to be a constant $\lambda$ and fertility is
taken to be a constant $f$ tuned to produce a
stationary population at equilibrium.  
We begin by imposing no upper age limit on fertility 
and later consider cases with fertility set back to zero
beyond an age $\beta < \infty $.

Our goal in the rest of this section is to rewrite 
the equilibrium equation for $h$
in a form that is easier to solve in special cases.  
We introduce notation for the indefinite integral 
of the aggregate population survivorship function \eqref{E:aggsurv}.
\begin{equation}
\label{E:Tofa}
T(a) := \int_a^{\infty} \, \exp \left( -\int_{\alpha}^x \, 
                    [ \lambda +  h(y)] dy \right)  \, dx 
\end{equation} 
This quantity $T(a)$, often written $T_a$,  is the same 
as the column for ``remaining person-years lived'' 
in the population lifetable.   Since for simplicity 
we have been omitting juvenile mortality,  
life expectancy $\xi$ at the age of maturity $\alpha$  
is the same as $T(\alpha)$ and equal to the reciprocal
of the level of constant fertility required for stationarity:  
\begin{equation}
\xi = T(\alpha)  = T(0) - \alpha = 1/f.  
\end{equation}

The equilibrium condition \eqref{E:kapintegral} can now
be written in terms of $h(a)$ and $T(a)$:
\begin{equation}
\label{E:solelyha}
\nu(a) = h(a) \int_a^{\infty} \, f  \, 
              \exp \left( -\int_{\alpha}^x \, ( \lambda +  h(y)) dy \right)  \, dx 
              =  h(a) \, f \, T(a) 
\end{equation} 
\par\indent
Since $ -T^{\prime}(a)$ is the survivorship function and
minus the derivative of log survivorship is the aggregate
hazard,  $ \lambda + h(a) $ 
is $ T^{\prime\prime}(a)/T^{\prime}(a)$.  
The equilibrium condition \eqref{E:kapintegral} 
therefore implies that $T(a)$ must satisfy 
a non-linear second-order differential equation: 
\begin{equation}
\label{E:ode}
\nu(a) = \left(  \frac{ T^{\prime\prime}(a)}{T^{\prime}(a) } 
               - \lambda \right) \, f \, T(a)     
\end{equation}

Solution of this differential equation is facilitated
by exploiting the monotonicity of $T(a)$ to change variables
from age $a$ to person-years $\tau = T(a)$.  
We make use of the inverse function $ T^{-1}$ 
which is defined to satisfy $ T^{-1}(T(a)) = a $
and $T( T^{-1}(\tau) ) = \tau $.  The symbol $\circ$ 
denotes composition of functions:  $ T^{-1}(T(a)) $
is the same as $ T^{-1} \circ T (a) $.  
Aggregate survivorship is expressed as a function 
of $\tau $ by composing the derivative of $T(a)$
with the inverse of $T(a)$:
\begin{equation}
L(\tau) \, := \, -T' \circ T^{-1} (\tau)
\end{equation} 
With this definition,  $ L(T(a)) = \mathbb{E}_{\rho} \, [ l_a (G) ]  $.
The function $L$ is easy to interpret.  The last $\tau$ 
person-years lived by members of a cohort are lived 
by the fraction $L$ of the members. 

The derivative of $L(\tau)$ with respect to $\tau$ comes 
out to be the hazard expressed as a function of $\tau$: 
\begin{equation}
L'(\tau) = \frac{T''\circ T^{-1}(\tau)}{-T'\circ T^{-1}(\tau)} 
               =  h \circ T^{-1}(\tau)  + \lambda
\end{equation}
From the definition of $\xi$, we have $ L(\xi) = 1$. 
The function $L$ must also satisfy the boundary 
condition $ L(0) = 0 $ with $ L(\tau) >  0$ for $\tau > 0$.
At an age to which no one survives, there are no 
remaining person-years to live.   

We can rewrite $ \nu = h f T $ in the form $ h = \nu/(fT)$,
substitute for $h$, and  
express both sides as functions of $\tau$: 
\begin{equation}
\label{E:Lprime}
L'(\tau) =  \lambda + \frac {\nu \circ T^{-1} (\tau)}{f \tau}  
\end{equation}

\section{
Unraveling}
\label{sec.unravel}

\par\indent

The simplest case of mutation accumulation 
with point-mass profiles takes the mutation 
rate $\nu$ to be a constant $\nu_0$ at all reproductive
ages.    

In the linear framework,  this case is the
starting-point for the notable results 
of \cite{bC01},   
discussed in Section \ref{sec.force}. 
With a constant background hazard at adult ages,
we have baseline survival, remaining person-years lived,
and Hamilton's force of natural selection all going 
down  exponentially with age.  With a constant 
mutation rate in the numerator and the force in the
denominator of the linear approximate formula,
Equation \eqref{E:hamha},  we have the mean intensity of 
mutations and the increment to the population hazard both
going up exponentially with age,  achieving a 
total hazard equal to an exponential plus a  constant,  
a Gompertz-Makeham form.

What impact do the non-linear interactions 
suppressed in Hamilton's formula turn out to 
have on this important prediction?  
We seek solutions to Equation \eqref{E:kapintegral} 
with $\nu(a) \equiv  \nu_0$.  
 
Plugging in a constant for $\nu$, the
solution to the differential equation \eqref{E:Lprime} 
including a constant of integration $A$ is given by
\begin{equation}
L(\tau) = \lambda \tau + (\nu_0/f ) \,  \log(\tau)  + A  
\end{equation}
We expect to determine the constant of integration
from the initial condition $ L(\xi) = l_{\alpha}  = l_0 = 1 $, 
since $T(\alpha) = \xi $.  We also expect to set $ f = 1/\xi $
to tune the population growth rate to stationary levels.  
A puzzle arises,  because $ \xi$, the life expectancy 
at maturity,  is as yet an unknown quantity that should
be determined by the equations, while it appears that 
different values of $\xi $ can correspond to different
choices for $A$.  But a deeper problem intervenes.  
The quantity $\log(\tau)$ goes to minus infinity as the 
remaining person-years of life $\tau$ go to zero,
forcing there to be some non-zero value of $\tau$ for
which $L(\tau)$ vanishes.   But our model requires
that $L(\tau) = 0 $ only when $\tau = 0 $.  
We cannot have further person-years to be lived 
when no survivors remain.  
This contradiction shows that the non-linear model has
no equilibrium solution in this elementary case.
This unexpected finding is the first of our main results.

We can visualize the disappearance of an equilibrium
in several ways, with respect to age, with respect to 
time, or with respect to the shape of the mutation
rate function.     

The picture with respect to age
has been mentioned in Section \ref{sec.linear}.  
We give a sketch rather than a formal argument. 
We seek to construct a solution satisfying the equilibrium 
condition \eqref{E:solelyha} restricted to some late range 
of ages $[z, \infty) $ and then seek to extend the
construction backward to earlier ages.
At late ages, selective pressure is driven to low levels by 
the background hazard as well as by any late-acting alleles. 
Beyond some sufficiently late $z$, we expect the pressure to 
be too weak to be balancing a rate of new mutations 
that does not drop with age. We therefore expect infinite $h$ 
and a Wall of Death.  At the next earlier ages,  the 
elimination of later reproduction by the Wall of Death 
again leaves weak selective pressure.  The interaction
between older and younger ages, missing from the 
linear approximate model,  puts $T(a)$ near zero. 
Low $T(a)$ requires high $h(a)$ to 
make   $ h(a) f T(a) $ as large as $\nu_0$ 
but high $h(a)$ implies even lower $T(a)$,
demanding, as it turns out, that the Wall of 
Death be earlier. Each Wall of Death implies
an earlier Wall of Death, the instability propagates 
down through the whole reproductive span,  and 
our construction unravels.

A complementary picture with respect to time 
is implied by the dynamic equation \eqref{E:rhot} 
with the null genotype as starting state.
A steady influx of mutations affects 
the whole reproductive span,  and $h_t(a)$ begins 
to increase over time like $\nu_0 \, t $ at all ages. 
At older ages, where selective pressure is always
low,  this linear increase continues unabated,
whereas at younger ages it is slowed for a while by
outflow due to natural selection maintained 
by substantial values of $ f T_t(a) $.  
The function $ h_t(a) $ at a snapshot in time
has an age profile which keeps low for a stretch of ages, 
climbs as $T_t(a)$ drops off with age,
and settles out at $ \nu_0 \, t $.
As time goes by, the climbing phase accompanying
the drop in $T_t(a)$ shifts down to younger and
younger ages,  until the hazard rate at every
adult age comes to be marching toward infinity.

Details of the dynamics depend on assumptions
about fertility.    We may hold fertility fixed
over time, but we have to recognize that  
no fixed fertility level is sufficient for stationary
population growth at equilibrium when there is no equilibrium.  
Unbounded accumulation of mutations across the 
whole reproductive span drives any population to extinction.
We may instead let fertility levels adjust over time 
to maintain stationarity with current values of $h_t(a)$,  
on an assumption that feedback between resources and
population growth operates on a faster timescale than
mutation and selection.  Under this scenario, the
climbing phase in the age profile of $h_t(a)$ steepens with 
age and time as it shifts to younger ages,  and the fertility
level heads toward infinity.   

A third way of picturing the disappearance of an equilibrium 
makes use of results from the next section about mutation
rate functions with different shapes.  Section \ref{sec.walls}
displays a family of examples in which the mutation rates 
are nearly constant but not exactly constant.  Each mutation 
rate function has a drop down to zero with some characteristic
steepness which turns out to produce an equilibrium 
with a Wall of Death.  The nearer the mutation rate function 
to constancy,  the nearer is the Wall of Death to the age
at maturity, the shorter the reproductive lifetime, 
and the higher the fertility level required for stationarity.  
In order to reach the case of a wholly constant mutation 
rate function the brief high burst of fertility before 
death would have to turn into a delta function or point mass 
at the age of maturity, followed by immediate death.     

Such a stylized life history,  called the ``salmon limit'' 
by \cite{sT97},   is a far cry from the smooth Gompertz-Makeham
equilibrium from the linear approximate model.   
Non-linear interactions make the Gompertz-Makeham form
collapse,  leaving no smooth equilibrium for the elementary
case of constant mutation rates with point-mass increments. 

This outcome depends on recombination.  
Recombination spreads 
the deleterious alleles throughout the
population,  leaving no lineages untouched
by a surfeit of late-acting mutant alleles.
In the absence of recombination,  as shown 
in \cite{SEW05}, a minority group of high-fitness
lineages can keep the aggregate population
hazard finite at younger ages in the face of constant 
mutation rates and a late-age Wall of Death.  

It is remarkable that the collapse of
the equilibrium in our model with Free Recombination
does not depend on the magnitude of the uniform 
mutation rate $\nu$.  Even the tiniest such rate 
cannot be balanced by the force of natural selection.   
The elementary case with constant mutation rates
and point-mass increments is the most studied 
case for the linear approximate model.
The transforming effect of non-linear interactions
for this case of all cases is a dramatic 
\it
denouement. 
\rm

\section{Walls of Death}
\label{sec.walls}

\par\indent

Unraveling can be avoided in several biologically 
reasonable ways, already adumbrated, for instance, 
in  \cite{kW97}, pages 11ff.
There may not be mutant alleles whose effects
are entirely concentrated at or above an age of onset.
An ``entry cost'' of some small loss of fitness at 
young ages associated with all later-acting mutations 
will keep unraveling in check.   Restricting ages of 
onset to some finite subset of discrete ages will also 
suffice.   Here we examine another alternative,
mutation rates for point-mass increments that
remain only nearly constant with age of onset
and drop to zero at what comes to be the
end of life.

We construct a family of cases with nearly constant 
mutation rates by seeking forms for the remaining 
person-years function $T(a)$ satisfying Equation \eqref{E:Lprime}
consistent with a predetermined relationship between $T(a)$ and $\nu(a)$.  
We then construct $\nu(a)$ to validate this relationship
and deduce $h(a)$ from $T(a)$.  Our cases are indexed
by a positive exponent $\theta$ less than or equal to $1$.   
Our intended relationship between $T(a)$ and $\nu(a)$ 
for $ a > \alpha $ takes the form 
\begin{equation}
\label{E:nuofTofa}
\nu(a)  =   \nu_0 ( \, T(a)/ T(\alpha)\, ) ^{\theta}
\end{equation}
When  $ \theta $ is close to zero,  the rate 
starts at $\nu(\alpha) = \nu_0 $ and remains 
nearly constant until close to the end of life.

With $\nu$ as in Equation \eqref{E:nuofTofa}, the 
differential equation \eqref{E:Lprime} takes the form 
\begin{equation}
L'(\tau) = \lambda +  (f \xi) ^{-1} \nu_0 (\tau/\xi)^{\theta - 1}
\end{equation}
Its solution is 
\begin{equation}
L(\tau) = \lambda \tau + (f \xi \theta)^{-1} \nu_0 \xi^{1-\theta} \tau^{\theta} + A
\end{equation}
Applying our boundary condition $L(0) = 0$ implies $  A = 0 $.
With $ f \xi = 1 $ for stationary
growth and $L(\xi) = 1 $ by the definition of $\xi$, life
expectancy at maturity is given by 
\begin{equation}
\label{E:mu}
\xi = \frac{ 1}{ \lambda + \nu_0/\theta }
\end{equation}
Higher mutation rates lower life expectancy, as they should,
and so does slower tapering of the mutation rates with lower $\theta$.
With no mutations,  $ \nu_0 = 0 $,  survival drops 
exponentially at the rate $\lambda$, and $\xi = 1/\lambda$.   

We can express $a$ as a function of $\tau$, because we 
now know the derivative of $a$ with respect to $\tau$,
the reciprocal of $-L(\tau)$.
\begin{equation}
\label{E:aforxi}
a  - \alpha =  \int_{T(\alpha)}^{T(a)} \frac{-d\tau}{L(\tau)}
           =  \int_{T(a)}^{\xi} \frac {d\tau}
     {\lambda \tau + (f \xi \theta)^{-1} \xi^{1-\theta} \,\nu_0 \, \tau^{\theta}}
\end{equation}

This expression can be integrated in closed form
and the answer inverted to give $\tau$ as a function of age.
We change notation for age from $a$ to $x$ for subsequent clarity:
\begin{equation}
\label{E:tauofx}
\tau(x) = \xi \left( ( 1 +  \frac{ \nu_0}{\lambda \theta} )
                     ( \exp(-\lambda(1-\theta)(x-\alpha)) ) 
              - \frac{ \nu_0}{\lambda \theta}  \right)^{1/(1-\theta)} 
\end{equation}
%
When we substitute into $L^{\prime}(\tau)$,  we find 
the contribution of genetic load to the hazard rate to be
\begin {equation}
\label{E:hazardxi}
h(x)  =  \frac{ \nu_0 \exp ( \lambda (1 - \theta) ( x -\alpha ) ) }
          { ( 1 + \nu_0/(\lambda \theta)) - 
           \nu_0/(\lambda \theta) \exp ( \lambda (1 - \theta) ( x - \alpha ) ) }
\end{equation} 

The hazard rate goes to infinity as $x$ 
approaches  
\begin{equation}
\label{E:omega}
\omega = \alpha +  \frac{ 1}{ \lambda(1 - \theta)} 
                 \log \left( 1 + \frac{ \lambda \theta}{\nu_0}  \right)  
\end{equation} 
For $ \theta < 1 $, $\omega $ is the age of a Wall of Death.

We now need to write down a mutation rate $\nu$ as a function of 
age which satisfies the posited relationship with $T$ 
and leads to \eqref{E:hazardxi} as an equilibrium solution.
The posited relationship follows easily 
by raising $ \tau(x)/\xi $ from \eqref{E:tauofx} 
to the power $\theta$.   However, this relationship does
not tell us how to define $\nu(x)$ beyond the age $\omega$
at which $\tau$ vanishes.  

It is tempting to define $ \nu(x) \equiv 0 $ beyond $\omega$, 
but if baseline fertility remains positive beyond $\omega$,
the resulting equilibrium will not be the limit over time of
the dynamical process starting from the null genotype.  
Late-age fertility will keep $h_t(a)$ bounded and prevent
the Wall of Death.  A second alternative is to terminate
the baseline reproductive span at an age $\beta$ equal 
to $\omega$.  As with the linear approximate model,
a limiting Wall of Death does then occur at the end
of reproduction.   

A third alternative is to set $\nu(x)$ back equal 
to some positive constant after its
drop to zero at $\omega$.  The rate is pinched to zero 
at $\omega$ rather than cut off.   The dynamical process
starting from the null genotype does then converge to 
the equilibrium given by \eqref{E:hazardxi}.    
This alternative is of theoretical interest, showing 
as it does that a Wall of Death can occur before rather than
at the last age of reproduction in the full non-linear model.

The behavior of our family of cases as the exponent $\theta$ 
approaches zero has been discussed in Section \ref{sec.unravel}.
The mutation rate becomes more and more nearly constant
and the Wall of Death at $\omega $ moves down 
to the age of maturity,  wiping out all fertility.  

For ages $x$ well below the Wall of Death,  the denominator
of Equation \eqref{E:hazardxi}  is nearly constant,  
and the hazard along with the constant baseline contribution
approximates a Gompertz-Makeham form.   As $x$ comes closer
to the Wall of Death,  the hazard function becomes hyperexponential. 
Derivatives of all orders go to infinity.   

The predicted hazards from the linear model with the same form
for $\nu$ have exponential increase at the same rate as the
numerator in Equation \eqref{E:hazardxi}.  If no upper age limit
is imposed on the reproductive span,  the Gompertz pattern
continues out to infinity.   If an upper age limit is imposed,
the linear model has a Wall of Death at that age, but 
not before.   

Mutation rates that drop sufficiently rapidly with 
ages of onset will prevent Walls of Death in the absence of
an upper limit to ages of reproduction.    Indeed,  
Equation \eqref{E:solelyha} allows us to start with a fairly
arbitrary target shape for $ h(a) $ and find a set of mutation
rates $\nu(a) $ which will generate it.   The case $\theta = 1 $
from our family of cases has an exponentially declining mutation
rate which produces an increment to the hazard rate that is
constant over age.

\section{Haldane's Principle }
\label{sec.haldane}

\par\indent

Our equilibrium condition \eqref{E:solelyha} encapsulates 
a notable result.  In our non-linear model with point-mass increments,  
if an equilibrium exists, the aggregate population hazard
maintained by mutation-selection balance at equilibrium does
not depend on the sizes $\eta(a)$ of the mutational effects.

This result generalizes a property 
of linear approximate models,  but in a surprising
direction.  Scaling up the size of the effect of a mutation
increases the selective pressure against it and reduces its
expected frequency.  In the linear setting,  the expected 
hazard, that is, the average hazard averaged across the population,
is insensitive to $\eta(a)$.  Doubling $\eta(a)$ halves the
expected frequency $\rho(a)$ and leaves the expected 
hazard $\eta(a)\rho(a)$ unchanged.  
In our full non-linear setting,  it is not the expected hazard
but the aggregate population hazard that comes out to be
invariant to changes in $\eta(a)$.  

The difference between aggregate and expected hazards is
a preoccupation of social scientists. 
It is due to demographic selection or culling, already
discussed in Section \ref{sec.linear}.  
The population is heterogeneous.  Some members have genotypes
with more mutant alleles and lower net reproduction. Others have
fewer mutant alleles and higher net reproduction.   The 
heterogeneity is what allows natural selection to operate over
time, and it is also what produces demographic selection with
increasing age. 
Members with more mutant alleles die at younger ages,
and survivors to later ages carry smaller than average 
sets of mutant alleles.   The aggregate population hazard,
based on proportions surviving, is lower at
advanced ages than the expected hazard,  which averages over
all members, irrespective of survival.      

In the linear framework, one way of calculating overall
selective costs makes them equal the total mutation
rate.  The contribution from a number $ \nu(a)/(\eta w(a))$
of mutant alleles acting at $a$, each with a cost $\eta w(a)$,
amounts to $\nu(a)$, and the integral over ages of onset
equals the total mutation rate.  However, calculating
selective cost directly from the predicted hazard 
gives a different answer, reflecting the inconsistencies
built into the linear approximations.   

In the full non-linear model,  the change in expected 
frequency $\rho(a)$ which compensates for a change in effect
size $\eta(a)$ is only approximately linear in $1/\eta(a)$.  
But the non-linearity of the change exactly balances the
non-linear effect of culling. 
Cases in which most population members carry 
only a handful of mutant alleles, each with a large effect,
lead to the same aggregate hazard as cases in which 
almost all population members carry a huge number of 
mutant alleles, each with a tiny effect.  
The variance in net reproduction across the heterogeneous
population is very different in the two settings,  but
the aggregate hazard,  the most readily observable outcome,
is the same.

The total mutation rate, obtained by integrating over $m$,
turns out to equal a function of the aggregate population 
survivorship $ \mathbb{E}_{\rho} \left[ l_x(G) \right] $.
The expectation is taken over the Poisson point-process 
distribution for $G$
determined by the equilibrium intensity $\rho$.  
\begin{equation}
\label{E:haldane}
\nu(M) = \int_{\alpha}^{\infty}  \,  
       ( - \log \left( \mathbb{E}_{\rho}\,[l_x(G)/l_x(0)]) \right) 
       \, f_x \, \mathbb{E}_{\rho} \,  [ l_x(G) ]  \, \,  dx 
\end{equation} 

The right-hand side is a fertility-weighted version of
lifetable entropy, described with references in the
textbook by \cite{KC05}, pages 80--82 and 166.
In a sense, this equation is a 
generalization of Haldane's Principle to 
our age-dependent models for mutation-selection balance under
Free Recombination with point-mass profiles. 
The quantity equal to the total mutation rate, however,
is not the overall loss of fitness for the population. 
The overall loss in fitness depends only on the aggregate
population hazard rate,  so it remains independent of
the sizes of mutational effects as Haldane posited,
but not entirely independent of the age pattern of
mutation rates. 

\section{Conjectures and Conclusions}

\par\indent

The derivations in this paper show that non-linear 
interactions can make a profound difference to 
patterns of senescent mortality produced from mutation
accumulation.    Breakdowns can occur not only when
rates of deleterious mutations are high but, in
the simplest  cases, whenever they are not zero.

The setting with point-mass mutations is the one 
most thoroughly studied in the past,
going back to Hamilton himself.   
It is a stylized setting.  Do our main 
conclusions hold when the age-specific
effects of deleterious mutations are 
not concentrated at single ages but
spread across a range?

We conjecture that unraveling
does occur not only with point-mass profiles but
with profiles from translation families 
of the kind also treated by \cite{bC01}. 
Let the functions $\kappa(m,x)$ each vanish
below some age of onset $a(m)$ and have the
same shape above it, being shifted versions
of some template.  We expect that a mutation
rate $\nu$ constant over ages of onset 
still leads to unraveling.  We also expect
that the shapes can vary to some extent
and that the strict absence of effects below
the ages of onset is the feature that drives
models to unravel.   With mutation
rates that taper with age,  such absence
leads to early Walls of Death.

A reasonable way for nature to avoid such 
pathologies would be for deleterious mutations 
to have small effects of at least some minimal 
level at young ages even when their main effects 
are concentrated later.   Such a requirement
would keep genetic loads from growing to infinity.   
In the process, it would introduce a tendency for hazard
rates to tend toward plateaus at high ages,
\cite{bC01} 
suggested this condition within the linear framework
as a way of generating plateaus.  With the non-linear
model, the argument becomes stronger.  The necessity
to avoid unraveling becomes a reason for expecting
plateaus.

We have seen that Gompertz-Makeham hazards arise quite
easily at young and medium ages from our non-linear
model with point-mass profiles and tapering mutation
rates.  We conjecture that they also arise 
with distributed mutational profiles of a more
realistic kind.  Applications of our model with
richer families of profiles will be presented in 
a sequel.

A number of aspects in the application of our
non-linear model await examination.  
These include 
\begin{itemize}
\item
possible closed-form solutions with point-mass
profiles augmented with fixed early-age selective
costs;
\item
inclusion of mutations depressing fertility as
well as augmenting hazards,  with point-mass
profiles and with more realistic profiles;
\item
extension of the results of \cite{aB08}
for effects that act multiplicatively on hazards,
exploiting the full non-linear model;
\item
comparisons of predicted hazards between our
model with Free Recombination and the alternative
model without recombination;
\item
study of intermediate assumptions about recombination
and their implications for unraveling and Walls of Death;
\end{itemize}

A larger goal is to begin to integrate the non-linear
models for mutation accumulation examined here with 
models for other contributors to senescent processes.
Mutation accumulation does not act in isolation.
It reshapes vital schedules that themselves reflect 
cellular and organismic processes and considerations
of life-history optimization in interactions with
environments.   Mathematical modeling of mutation
accumulation is a point of departure for further enhancements
of our evolutionary understanding of senescence.  

\newpage


\end{document}